\documentclass[12pt]{article}

\usepackage{latexsym}
\usepackage{amsfonts}

                   {\hfil\hfill$\square$\endtrivlist}

\begin{document}

\title{New conservation laws for electromagnetic fields in gravity}

\author{J M M Senovilla\footnote{Dedicated to Professor Jos\'e A. Azc\'arraga 
on the occasion of his 60th birthday.}\\
Departamento de F\'{\i}sica Te\'orica, Universidad del Pa\'{\i}s Vasco,\\
Apartado 644, 48080 Bilbao, Spain \\
e-mail: \tt{wtpmasej@lg.ehu.es}}

\maketitle

\begin{abstract}
A recently found \cite{BES} 2-index, symmetric, trace-free, {\em divergence-free}
tensor is introduced for arbitrary source-free electromagnetic fields. The 
tensor can be constructed for any test Maxwell field in Einstein 
spaces (including proper vacuum), and more importantly for any Einstein-Maxwell 
spacetime. The tensor is explicitly given and analyzed in some special situations, 
such as general null electromagnetic fields, Reissner-Nordstr\"om 
solution, or classical electrodynamics. We present an explicit example 
where the conserved currents derived from the energy-momentum tensor 
using symmetries are trivial, but those derived from the new tensor are not.

\end{abstract}

\section{Introduction: electromagnetic fields}
Let $(V,g)$ be a 4-dimensional spacetime with metric $g$ 
(signature -,+,+,+). An electromagnetic field is any 2-form 
\mbox{\boldmath $F$} satisfying the Maxwell equations (e.g. \cite{LL,MTW})
\begin{equation}
    d\mbox{\boldmath $F$}=0\, , \hspace{5mm} 
    \delta \mbox{\boldmath $F$}= -\mbox{\boldmath $j$}\hspace{1cm}
\Longleftrightarrow \hspace{1cm} \nabla_{[\mu}F_{\nu\rho]}=0\, , \hspace{5mm}  
\nabla_{\sigma}F^{\sigma\nu}=j^{\nu} \label{maxeq}
\end{equation}
where $\vec j$ is the electromagnetic current 4-vector. As usual, one 
can define the dual of \mbox{\boldmath $F$} and the associated complex 
self-dual 2-form \mbox{\boldmath $\cal{F}$} by means of
$$
\stackrel{*}{F}_{\mu\nu}\equiv \frac{1}{2} 
\eta_{\mu\nu\rho\tau}F^{\rho\tau} \, , \hspace{15mm} 
\mbox{\boldmath $\cal{F}$}\equiv 
\frac{1}{2}(\mbox{\boldmath $F$}-i \stackrel{*}{\mbox{\boldmath 
${F}$}})
$$
where \mbox{\boldmath $\eta$} is the canonical volument element 
4-form. A duality rotation (e.g. \cite{PR}) is any transformation of the type
$$
\mbox{\boldmath $\cal{F}$}'=e^{i\theta}\mbox{\boldmath $\cal{F}$} .
$$
The {\em source-free} Maxwell equations---that is, (\ref{maxeq}) 
with $\vec j =\vec 0$---, which can be written simply as 
$\nabla_{\sigma}{\cal F}^{\sigma\nu}=0$, remain clearly invariant 
under duality rotations. 

The energy-momentum tensor of \mbox{\boldmath $F$} can be written in 
any of the following forms
\begin{equation}
 T_{\mu\nu}=F_{\mu\rho}F_{\nu}{}^{\rho}-
\frac{1}{4}g_{\mu\nu}F_{\rho\sigma}F^{\rho\sigma}
=\frac{1}{2}\left(F_{\mu\rho}F_{\nu}{}^{\rho}+
\stackrel{*}{F}_{\mu\rho}\stackrel{*}{F}_{\nu}{}^{\rho}\right)
=2{\cal F}_{\mu\rho}\overline{{\cal F}}_{\nu}{}^{\rho} \hspace{1cm}  
\label{emt}
\end{equation}
where an overbar denotes complex conjugation. Obviously, $T_{\mu\nu}$ 
is symmetric, traceless, and invariant against duality rotations. 
Furthermore, it satisfies the {\em dominant energy condition}, that is to say
$$
T_{00}\geq |T_{\mu\nu}| \,\, \mbox{in all orthonormal bases} 
$$
or equivalently, see e.g. \cite{HE,S}
$$
T(\vec u,\vec v)\geq 0
$$
for all future pointing (ergo non-spacelike) vectors $\vec u$ and $\vec 
v$. Hence, if the energy density with respect to an observer vanishes, 
then the whole energy-momentum tensor is zero, and the 
electromagnetic field is absent. Another important relation satisfied 
by this tensor is the algebraic Rainich identity 
\cite{Rai,MW,PR,Exact,BS}
\begin{equation}
T_{\mu\rho}T^{\rho}{}_{\nu}= 
\frac{1}{4} (T_{\sigma\rho}T^{\sigma\rho}) g_{\mu\nu} \label{rainich}
\end{equation}
which fully characterizes the explicit form (\ref{emt}) if 
$T^{\mu}{}_{\mu}=0$ is taken into account.

A trivial calculation using 
(\ref{maxeq}) and (\ref{emt}) leads to
\begin{equation}
    \nabla_{\mu}T^{\mu\nu}=F^{\nu\rho}j_{\rho} \,\,\,\, \mbox{whence:}
 \,\,\,\, \vec j =\vec 0 \, \Longrightarrow \,\, 
 \nabla_{\mu}T^{\mu\nu}=0\, .\label{emcon}
\end{equation}
This encodes the ``covariant'' conservation of energy and momentum 
for electromagnetic fields in the absence of charges and electric 
currents. It also implies that, for any {\em conformal Killing 
vector} $\vec\xi$ \cite{Exact}, the corresponding Killing current 
$\vec{{\cal J}}$ is divergence-free:
\begin{equation}
{\cal J}^{\mu}(\vec\xi)\equiv T^{\mu\nu}\xi_{\nu} \, \Rightarrow \,\,
\nabla_{\mu} {\cal J}^{\mu}=0 \label{jem}
\end{equation}
providing conserved quantities via Gauss' theorem \cite{HE,MTW}, see e.g. 
\cite{LSV,S} for some explicit examples. 
On the other hand, if there are electric charges and currents on the 
spacetime, one can still restore conservation of the {\em total} 
energy and momentum by adding the corresponding energy-momentum 
tensors of those sources. A typical example can be found in 
\cite{LL} for classical electrodynamics.

\section{The main result: $H_{\mu\nu}$}
For any electromagnetic field, there exists another tensor quadratic 
in \mbox{\boldmath $F$} and with the same 
properties as $T_{\mu\nu}$ except for the dominant energy condition. 
This tensor can be written in any of the following equivalent forms
\begin{eqnarray}
  H_{\mu\nu}=\nabla_{\tau}F_{\mu\rho}\nabla^{\tau}F_{\nu}{}^{\rho}-
\frac{1}{4}g_{\mu\nu}\nabla_{\tau}F_{\rho\sigma}\nabla^{\tau}F^{\rho\sigma}=
\hspace{4cm}\nonumber\\
=\frac{1}{2}\left(\nabla_{\tau}F_{\mu\rho}\nabla^{\tau}F_{\nu}{}^{\rho}+
\nabla_{\tau}\stackrel{*}{F}_{\mu\rho}
\nabla^{\tau}\stackrel{*}{F}_{\nu}{}^{\rho}\right) 
=2\, \nabla_{\tau}{\cal F}_{\mu\rho}\nabla^{\tau}\,
\overline{{\cal F}}_{\nu}{}^{\rho}   
\label{H} 
\end{eqnarray}
from where immediately follows that $H_{\mu\nu}$ is symmetric, 
traceless, and invariant under duality rotations (with constant 
$\theta$ !). More importantly, it is also divergence-free if the 
source-free Maxwell equations hold
\begin{equation}
    \nabla_{\mu}H^{\mu\nu}=\nabla^{\tau}F^{\nu\rho}\nabla_{\tau}j_{\rho} 
    \,\,\,\, \mbox{whence:} \,\,\,\,
 \vec j =\vec 0 \, \Longrightarrow \,\, \nabla_{\mu}H^{\mu\nu}=0\, .\label{Hcon}   
\end{equation}
This is valid (i) if the electromagnetic field is ``test'', that is to 
say, it does not contribute to the righthand side of the Einstein 
field equations, so that the spacetime is vacuum with a possible 
cosmological constant. And (ii) in the full non-linear
theory such that the Einstein-Maxwell field equations with a possible 
cosmological constant are 
satisfied. The proof of these results is far from obvious using tensor 
calculations. Actually, its derivation was achieved using spinors in 
\cite{BES}. For a full proof involving only tensor techniques one can 
consult \cite{EW} (see also \cite{eh2002}, where related results for 
a ``Lanczos potential'' $L_{\mu\nu\tau}$ in place of 
$\nabla_{\tau}F_{\mu\nu}$ were found).

Analogously to the case of $T_{\mu\nu}$, using (\ref{Hcon}) one can construct 
divergence-free vector fields associated to any conformal Killing 
vector $\vec\xi$ by means of
\begin{equation}
J^{\mu}(\vec\xi)\equiv H^{\mu\nu}\xi_{\nu} \, \Rightarrow \,\,
\nabla_{\mu} J^{\mu}=0 \label{jH}
\end{equation}
providing new conserved quantities in any Einstein-Maxwell spacetime (or any 
Einstein space with a test \mbox{\boldmath $F$}) having such $\vec\xi$.

Note finally that, generically, the dominant property
(i.e. that $H(\vec u,\vec v)\geq 0$ for all future-pointing $\vec 
u,\vec v$) need not be satisfied.

An obvious question arises: why was $H_{\mu\nu}$ unknown until very 
recently? One reason may be, as previously stated, that 
the proof of (\ref{Hcon}) involves difficult calculations. Another 
one may be the circumstance that its physical units are not those of 
energy density. Actually, $H_{\mu\nu}$ has units of ``superenergy'' 
in analogy with the 
well-known Bel-Robinson tensor, see e.g. 
\cite{Bel1,Bel2,Bel3,chevreton,PR,S}, which means units of energy 
density per unit surface.
One can further prove (see \cite{BES}) that $H_{\mu\nu}$ in the unique trace 
of the 4-index superenergy tensor of the electromagnetic field, as 
given originally by Chevreton \cite{chevreton} or in the unified 
general treatment of \cite{S,S1}. Thus, if the new tensor has any physical
relevance at all, this could help in understanding the concept of 
gravitational superenergy as well.

\section{Uniqueness of $H_{\mu\nu}$}
The tensor (\ref{H}) is unique and solely constructable for the source-free
electromagnetic field. I would like to stress these properties by 
presenting a list of more detailed points:

\paragraph
1 In Special Relativity, that is to say, in flat spacetime, 
there exist many tensors which are quadratic in \mbox{\boldmath $F$} 
and with the previous properties. One example is the so-called 
zilch tensor \cite{lip,kib}, but there are many other. For a full list, 
see \cite{AP,FN}. Several examples pertinent here are
$$
H^{(n)}_{\mu\nu}=\nabla_{\tau_{1}}\dots \nabla_{\tau_{n}}F_{\mu\rho}
\nabla^{\tau_{1}}\dots \nabla^{\tau_{n}}F_{\nu}{}^{\rho}-
\frac{1}{4}g_{\mu\nu}\nabla_{\tau_{1}}\dots \nabla_{\tau_{n}}F_{\rho\sigma}
\nabla^{\tau_{1}}\dots \nabla^{\tau_{n}}F^{\rho\sigma}
$$
for any natural number $n$. Observe that $H^{(0)}_{\mu\nu}=T_{\mu\nu}$ and 
$H^{(1)}_{\mu\nu}=H_{\mu\nu}$. Obviously, these tensors are traceless and 
symmetric, and furthermore they are divergence-free for any $n$ if the 
electromagnetic field has no sources. This is obvious from 
the commutation of the covariant derivatives in flat spacetime. 
However, {\em the only tensors of the previous list which keep the 
divergence-free property in curved spacetimes are $T_{\mu\nu}$ and $H_{\mu\nu}$}.

\paragraph
2 The divergence-free property of $H_{\mu\nu}$ in generic spacetimes
holds {\em in four dimensions exclusively}. Thus, (\ref{Hcon}) is not 
valid, {\em not even for test electromagnetic fields}, in higher dimensions.

\paragraph
3 The construction of $H_{\mu\nu}$ is not translatable to other 
physical fields. Take, for instance, a minimally coupled massless scalar 
field $\phi$. The tensor analogous to $H_{\mu\nu}$ can be seen to be
\begin{equation}
D_{\mu\nu}=\nabla_{\rho}\nabla_{\mu}\phi\, \nabla^{\rho}\nabla_{\nu}\phi
-\frac{1}{2}g_{\mu\nu}
\nabla_{\rho}\nabla_{\sigma}\phi\, \nabla^{\rho}\nabla^{\sigma}\phi .
\label{D}
\end{equation}
This can be derived in several independent ways, such as by taking the 
trace of the 4-index superenergy tensor of the scalar field 
\cite{S,S1,Tey}, or in direct analogy with the expressions (\ref{H}) 
and (\ref{emt}) if one recalls the energy-momentum tensor of $\phi$: 
$T_{\mu\nu}=\nabla_{\mu}\phi\nabla_{\nu}\phi
-\frac{1}{2}g_{\mu\nu}\nabla_{\sigma}\phi\nabla^{\sigma}\phi$. 
Nevertheless, in non-flat spacetimes, and despite assuming the 
massless Klein-Gordon equation $\mbox{\Large{$\Box$}}\phi =0$, 
{\em $D_{\mu\nu}$ is not divergence-free}. This happens even for test 
scalar fields, that is to say, in Ricci-flat spacetimes.

\paragraph
4 The existence of $H_{\mu\nu}$ in the {\em full Einstein-Maxwell theory},
specially as given in its form (\ref{H}) not 
involving any curvature terms---just first derivatives of \mbox{\boldmath $F$}--
may seem a little surprising.
For source-free {\em test} electromagnetic fields in Einstein 
spaces ---including the Ricci-flat case---, the existence of 
$H_{\mu\nu}$ can be understood as a ``second derivative'' of the 
energy-momentum tensor. To see this, let us start with the simplest 
case of flat spacetimes, where clearly
$2H_{\mu\nu}=\mbox{\Large{$\Box$}} T_{\mu\nu}$---if the source-free Maxwell 
equations (\ref{maxeq}) hold---due to the commutativity of the covariant 
derivatives. In curved spacetimes things are not so simple.
However, {\em in Einstein spaces}, there is a D'Alembertian-like operator 
$\mbox{\Large{$\Box$}}_{L}$ introduced by Lichnerowicz \cite{L} which commutes 
with the divergence. Thus, after some manipulations and 
using some tensor identities one can actually prove that 
$2H_{\mu\nu}=\mbox{\Large{$\Box$}}_{L}T_{\mu\nu}$ in Einstein 
spaces. This result has been pointed out in \cite{D} for the 
Ricci-flat case. 

One can thus ask whether a similar but more 
involved calculation permits a relation between the energy-momentum 
tensor and the new tensor via a second order operator {\em in the full 
non-linear Einstein-Maxwell case} (allowing for an additional cosmological 
constant). This has been solved by Edgar \cite{E} recently in an elegant 
manner. He shows that one can define in general an operator acting on 
divergence-free rank-2 tensor fields such that the result is 
divergence-free too. Nonetheless, this operator involves {\em curvature 
terms} in addition to second order derivatives. Thus, the path to 
prove the relation of this with $H_{\mu\nu}$ is still complicated. As 
a matter of fact, in order to rederive the expression (\ref{H}) one 
has to (i) restrict the discussion to 4 dimensions, (ii) use the 
Einstein field equations, (iii) then utilize the 
source-free Maxwell equations (\ref{maxeq}) explicitly, (iv) take 
into account the special Rainich identity (\ref{rainich}), and (v) 
manipulate the result employing a 
4-dimensional identity \cite{eh2002,EW} involving the curvature tensor and the 
electromagnetic field\footnote{Points (i) to (iv) 
could be analogously followed for the scalar field without 
problems---substituting the Klein-Gordon equation for the Maxwell equations---, 
because there is an identity similar to (\ref{rainich}) in that case 
too, see e.g.\cite{BS}. 
However, the final point (v) has no translation to the scalar field, and 
thus an expression without curvature terms, such as (\ref{D}), is 
impossible for divergence-free tensors in this case. 
This holds also for test scalar fields, as 
remarked in point 3.}. 

One could argue, as for 
instance in \cite{D}, that this deprives $H_{\mu\nu}$ of relevant ``independent
content'' from (\ref{emt}). This is not clear to me, though, for 
there are cases in which the divergence-free currents (\ref{jem}) 
constructed from the energy-momentum tensor are trivial while those 
built from the new tensor, as in (\ref{jH}), are not. An explicit 
example is presented in \S \ref{rob-tra}. The point here is that the 
commutativity of the divergence and the second order operator does not 
immediatey translates to the currents, as they involve (conformal) 
Killing vectors. Thus, the curved space derivatives of the 
infinitesimal symmetry are also involved here.

\section{Applications and explicit examples}
In this final section we present some explicit expressions for the 
new tensor $H_{\mu\nu}$ in some cases of physical interest. The 
results may shed some light into its interpretation and its 
potential usefulness.

\subsection{Null electromagnetic fields}
Null electromagnetic fields can be characterized by any one of the 
following equivalent conditions:
\begin{eqnarray}
{\cal F}_{\mu\nu}{\cal F}^{\mu\nu}=0 \,\,\, \mbox{or equivalently} \,\,\,
F_{\mu\nu}F^{\mu\nu}=\stackrel{*}{F}_{\mu\nu}F^{\mu\nu}=0, \nonumber \\
\exists \vec{\ell} : \,\,\, 
\ell^{\mu}F_{\mu\nu}=\ell^{\mu}\stackrel{*}{F}_{\mu\nu}=0 \hspace{1cm}
\mbox{(then necessarily $\ell^{\mu}\ell_{\mu}=0$)},\nonumber \\
\mbox{\boldmath $F$}=\mbox{\boldmath $\ell$}\wedge \mbox{\boldmath $p$}, 
\hspace{1cm} \stackrel{*}{\mbox{\boldmath $F$}}=
\mbox{\boldmath $\ell$}\wedge \mbox{\boldmath 
$q$}\hspace{2cm}\label{null}
\end{eqnarray}
where the unit spacelike vectors $\vec p$ and $\vec q$ are orthogonal 
to each other and orthogonal to $\vec \ell$, the latter being 
intrinsically defined (up to a proportionality factor)
by the null field and sometimes called ``wave 
vector''---it defines the null direction of propagation of the 
electromagnetic field---. It must be noted that $\vec p$ and $\vec q$ 
are not intrinsically defined by \mbox{\boldmath $F$} and are subject 
to redefinitions of the form $\vec p\rightarrow \vec p +a\vec\ell$,
$\vec q\rightarrow \vec q +b\vec\ell$ for arbitrary functions $a$ and 
$b$. In other words, only the direction of the wave vector $\vec\ell$ 
is intrinsically defined and therefore, its orthogonal spacelike 2-plane, 
which is spanned by $\vec p$ and $\vec q$ and sometimes called ``polarization 
plane'', is also intrinsic to the null \mbox{\boldmath $F$}.

Due to the above remarks, the energy momentum tensor should 
only depend on the wave propagation direction, and this is certainly 
so, as is well known, because using (\ref{emt}) and (\ref{null}) one 
immediately gets
$$
T_{\mu\nu}=\ell_{\mu}\ell_{\nu}\, .
$$
Hence, the full Einstein-Maxwell system of equations is given by ($R_{\mu\nu}$ 
is the Ricci tensor, units with $8\pi G=c=1$,
and the cosmological constant $\Lambda$ is kept) 
$$
R_{\mu\nu}=\ell_{\mu}\ell_{\nu} + \Lambda g_{\mu\nu}
$$ 
together with (\ref{maxeq}) appropriately specialized on using 
(\ref{null}). An old result due to Mariot and Robinson \cite{M,R} 
states that this last part requires a null vector field 
$\vec\ell$ which is geodesic and shear-free (see e.g. \cite{Exact} for these 
definitions), i.e.
$$
\ell^{\mu}\nabla_{\mu}\ell^{\nu} \propto \ell^{\nu}, \hspace{1cm}
\nabla_{(\mu}\ell_{\nu)}\nabla^{\mu}\ell^{\nu}=
\frac{1}{2}(\nabla_{\mu}\ell^{\mu})^2 \, .
$$
But then a series of results eventually summarized in a 
paper by Goldberg and Sachs (\cite{GS} and references therein) imply 
that actually the spacetime must be ``algebraically special'' in the 
sense that the Weyl tensor $C^{\alpha}{}_{\beta\mu\nu}$ has a repeated 
principal direction given by $\vec\ell$, see e.g. \cite{Exact} p.\ 
88, that is to say
$$
\ell^{\beta}\ell^{\mu}\, C^{\alpha}{}_{\beta\mu[\nu}\ell_{\tau]}=0\, .
$$
For the Petrov classification and the nomenclature about principal 
null directions, see e.g. \cite{Bel3,PR,Exact}. Using the 
Newman-Penrose notation (e.g.\ \cite{PR,Exact}) these can be written as 
$$
\Psi_{0}=\Psi_{1}=0, \hspace{1cm} \sigma= \kappa=0 
$$
in a null tetrad with $\vec\ell$ as first null vector. The first two 
here express that $\vec\ell$ is a multiple principal null direction of 
the Weyl tensor, and the 
second pair that it is geodesic and shear-free.

Let us turn our attention to $H_{\mu\nu}$ for these null 
fields\footnote{An analysis of the 4-index Chevreton tensor for null 
electromagnetic fields can be found in \cite{WZ}.}. If one plugs 
(\ref{null}) into (\ref{H}), a na\"\i ve calculation leads to 
expressions involving derivatives of $\vec p$ and/or $\vec q$. But 
this is undesirable, as we know that only $\vec\ell$ is intrinsically 
defined, so that an expression involving only $\vec\ell$ must exist. 
This can certainly be achieved, after a very long and not easy 
calculation where all the previous results and other identitites must be used. 
The result is
\begin{equation}
H_{\mu\nu}=\nabla_{\rho}\left[\ell_{(\mu}\nabla^{\rho}\ell_{\nu)}- 
\ell^{\rho}\nabla_{(\mu}\ell_{\nu)}-\ell_{(\mu}\nabla_{\nu)}\ell^{\rho}\right] .
\label{nullH}
\end{equation}
This is the general expression of $H_{\mu\nu}$ for null 
electromagnetic fields. Several alternative but equivalent formulae 
can be given, such as
$$
H_{\mu\nu}=\frac{1}{2}\mbox{\Large{$\Box$}} (\ell_{\mu}\ell_{\nu})+
\ell^{\alpha}\ell^{\beta} C_{\alpha\mu\beta\nu}-
\frac{4}{3}\Lambda\, \ell_{\mu}\ell_{\nu}
$$
which may also be useful in some particular situations. Nonetheless, 
this last expression depends explicitly on the curvature of the 
spacetime, and thus (\ref{nullH}) is preferable in principle, and 
physically more interesting.

Some particular relevant examples of null electromagnetic fields are 
examined next.
\subsubsection{Maxwell pp-waves}
The plane-fronted waves with parallel rays (pp-waves in short) are 
defined as those spacetimes containing a constant null vector field 
$\vec\ell$, see e.g. \cite{Exact} and references therein, that is 
$$
\nabla_{\mu}\ell_{\nu}=0 \, .
$$
They contain a large subclass of solutions depending on two arbitrary 
functions which satisfy the Einstein-Maxwell equations. As is obvious 
from the previous formula and (\ref{nullH}) one has for all these 
solutions $H_{\mu\nu}=0$. Thus, the new tensor is identically zero 
for the usual plane waves, including here the classical case of plane 
waves in Special Relativity.

\subsubsection{Robinson-Trautman type-D solution for a null Maxwell field}
\label{rob-tra}
In 1962, Robinson and Trautman \cite{RT} found, among many other algebraically 
special solutions, a particular spacetime which solves the 
Einstein-Maxwell system with a possible $\Lambda$ for a null 
electromagnetic field. The solution is of 
Petrov type D and has a 3-parameter group of isometries acting 
transitively on spacelike 2-surfaces. In local coordinates 
$\{u,r,x,y\}$, the line-element takes the explicit form
$$
ds^2= 
r^2(dx^2+dy^2)-2dudr+\left(\frac{2m(u)}{r}+\frac{\Lambda}{3}r^2\right)du^2
$$
where $m(u)$---the ``mass''--- is an arbitrary but non-increasing function of the 
retarded time $u$: $\dot{m}\leq 0$ (this solution is reminiscent of 
the more popular Vaidya's spherically symmetric radiating solution). 
The null electromagnetic field is given (up to duality rotations) by
$$
\mbox{\boldmath $F$}=\sqrt{-2\dot{m}}\, du\wedge dx
$$
and its wave vector field is
$$
\mbox{\boldmath $\ell$}=\frac{\sqrt{-2\dot{m}}}{r}\, du
$$
so that the only non-zero component of the energy-momemtum tensor is
$$
T_{uu}=-\frac{2\dot{m}}{r^2} \, .
$$
Using (\ref{nullH}), a direct computation leads in this spacetime to 
$$
H^{\mu}{}_{\nu}=\hat{H}^{\mu}{}_{\nu}+\left(\frac{\ddot{m}}{\dot{m}r}
-\frac{2m}{r^3}-\frac{4}{3}\Lambda\right)\, \ell^{\mu}\ell_{\nu}
$$
where $\hat{H}^{\mu}{}_{\nu}$ takes the simple form
$$
\left(\hat{H}^{\mu}{}_{\nu}\right)=-\frac{2\dot{m}}{r^4}\times
\mbox{diag}(-1,-1,1,1)
$$
which is a structure analogous to that of typical ``Coulombian'' 
fields (see subsection \ref{nonnull} below). Thus, the form of 
$H_{\mu\nu}$ splits as the sum of a ``Coulombian-like'' term, due to a 
``charge'' represented by $\sqrt{-2\dot{m}}$ (the ``mass loss'' or 
radiation rate) plus a typical null term with contributions from $m$, 
$\ddot{m}/\dot{m}$ and the cosmological constant.

Using now the three Killing vectors 
$$
\vec{\xi}_{i}=\{\partial_{x},\partial_{y},y\partial_{x}-x\partial_{y}\}
$$ 
one can construct conserved 
currents in the way given in (\ref{jem}) and (\ref{jH}). It is 
remarkable that, as is obvious, all the currents (\ref{jem}) constructed from the 
energy-momentum tensor are identically vanishing, 
$$
{\cal J}^{\mu}(\vec{\xi}_{i})=0,
$$
whence the corresponding conserved quantities are trivial. On the other hand, 
the divergence-free currents (\ref{jH}) constructed from the new 
tensor are non-vanishing in general. For any one of the three 
previous $\vec{\xi}_{i}$ one has
$$
\vec{J}(\vec{\xi}_{i})=-\frac{2\dot{m}}{r^4}\vec{\xi}_{i}\,\, , 
$$
leading to non-trivial conserved quantities, as can be easily 
checked. Thus, in this particular spacetime, the new tensor provides 
useful information, and conservation laws, involving the physically 
relevant magnitude $\dot{m}$.

\subsection{Non-null electromagnetic fields}
\label{nonnull}
In this case, it is not possible to give a general formula such as 
(\ref{nullH}) because there are two null directions defined by 
\mbox{\boldmath $F$}, and they need not be shear-free nor multiple 
null directions of the Weyl tensor. We consider briefly two simple examples.

\subsubsection{The Reissner-Nordstr\"om solution}
The more general spherically symmetric solution of the Einstein-Maxwell 
equations is given by the following line-element in
typical spherical coordinates $\{t,r,\theta,\varphi\}$
$$
ds^2=-\left(1-\frac{2m}{r}+\frac{q^2}{r^2}\right)dt^2+
\left(1-\frac{2m}{r}+\frac{q^2}{r^2}\right)^{-1}dr^2+
r^2\left(d\theta^2+\sin^2\theta\ d \varphi^2\right)
$$
where $m$ and $q$ are arbitrary constants representing the total
mass and electric charge of the body creating the gravitational field. 
Only the exterior asymptotically flat region defined by 
$1-2m/r+q^2/r^2>0$ will be considered. This
restricts the range of $r$ to $r>0$ if $m^2<q^2$, or to
$r>r_+\equiv m+\sqrt{m^2-q^2}$ if $m^2\geq q^2$. 
Then, the metric is static ($\partial_{t}$
is a hypersurface-orthogonal timelike Killing vector)
and spherically symmetric.

The electromagnetic field is given by 
$$
\mbox{\boldmath $F$}=\frac{q}{r^2}\,  dt\wedge dr
$$
so that its energy-momentum tensor in the coordinate basis is
$$
\left(T^{\mu}{}_{\nu}\right)=
\frac{q^2}{r^4}\times\mbox{diag}(-1,-1,1,1).
$$

A straightforward calculation leads to
$$
\left(H^{\mu}{}_{\nu}\right)=\frac{q^2}{r^6}
\left(1-\frac{2m}{r}+\frac{q^2}{r^2}\right)\times\mbox{diag}(-3,-1,2,2)\, .
$$
Notice the physical units. This particular tensor satisfies the 
dominant property as well. The conserved currents that can be 
constructed from here are similar to that of the energy-momentum 
tensor, see also \cite{LSV}.

\subsubsection{The ``Bertotti-Robinson'' solution}
As early as 1917, Levi-Civita \cite{LC} gave the only Einstein-Maxwell spacetime 
which is homogeneous with homogeneous electromagnetic field. However, 
this line-element is commonly known as the ``Bertotti-Robinson'' solution, 
as it was rediscovered in \cite{R2} and, with the cosmological 
constant, in \cite{Be}. Locally, it can be given in adequate 
coordinates by
$$
ds^2=-dt^2+dx^2+\cos^2(ax)dy^2+\cos^2(at)dz^2
$$
and the electromagnetic field is
$$
\mbox{\boldmath $F$}=\sqrt{2}a \cos(at) dt\wedge dz \hspace{5mm} 
\Longrightarrow \hspace{5mm}  \nabla \mbox{\boldmath $F$} =0
$$
so that $H_{\mu\nu}$ is identically zero. The energy-momentum tensor 
is in this case
$$
\left(T^{\mu}{}_{\nu}\right)=2a^2\times\mbox{diag}(-1,-1,1,1).
$$

\subsection{Classical electrodynamics}
In flat spacetime of Special Relativity, and using an inertial reference system,
one can define the associated electric $\vec E$ and magnetic $\vec B$ 
fields. Then, using dots for time derivatives and ${}_{,i}$ for 
derivatives with respect to the spacelike rectilinear coordinates $x^i$, 
one can obtain (0 is the time-component and $\cdot$ is the usual scalar 
product of 4-vectors)
\begin{eqnarray*}
H_{00}&=&\frac{1}{2}\left[\dot{\vec{E}}\cdot\dot{\vec{E}}+
\vec{E}_{,i}\cdot\vec{E}^{,i}+\dot{\vec{B}}\cdot\dot{\vec{B}}
+\vec{H}_{,i}\cdot\vec{H}^{,i}\right]=
\frac{1}{2}\mbox{\Large{$\Box$}} (\vec{E}\cdot\vec{E} +\vec{B}\cdot\vec{B})\\
H_{0i}&=&\dot{\vec{E}}\times\dot{\vec{B}}+\vec{E}_{,i}\times\vec{B}^{,i}=
\mbox{\Large{$\Box$}} (\vec{E}\times\vec{B})\\
H_{ij}&=&\frac{1}{2}\mbox{\Large{$\Box$}} T_{ij} .
\end{eqnarray*}

\subsection*{Acknowledgements}
This contribution is partly based on a collaboration with 
G. Bergqvist and I. Eriksson. I am very grateful to S Brian Edgar 
for many comments and for a draft \cite{E} containing interesting 
results relevant here. Financial support under
grants BFM2000-0018 of the Spanish CICyT and 
9/UPV 00172.310-14456/2002 of the University of the Basque 
Country is acknowledged.

\end{document}